\documentclass[12pt, a4paper]{article}
\pdfoutput=1

\usepackage{amsmath,amssymb,exscale,xcolor}
\usepackage{psfrag,graphicx}
\usepackage{enumerate}
\usepackage[colorlinks=True, citecolor=blue, linkcolor=blue, urlcolor=black]{hyperref}
\input epsf

\newcommand{\be}{\begin{equation}}
\newcommand{\ee}{\end{equation}}
\newcommand{\bea}{\begin{eqnarray}}
\newcommand{\eea}{\end{eqnarray}}

\newcommand{\eq}[1]{Eq.~(\ref{#1})}
\newcommand{\zir}{z_{\rm IR}}

\newcommand{\zc}{z_c}

\newcommand{\zuv}{z_{\rm UV}}
\newcommand{\f}{\frac}
\newcommand{\diag}{\operatorname{diag}}

\def\gappeq{\mathrel{\rlap {\raise.5ex\hbox{$>$}}
{\lower.5ex\hbox{$\sim$}}}}
\def\lappeq{\mathrel{\rlap{\raise.5ex\hbox{$<$}}
{\lower.5ex\hbox{$\sim$}}}}
\newcommand{\Tr}{\mathop{\rm Tr}}
\def\I{\rm 1\kern-.24em l}  

\begin{document}
\pagestyle{empty}
\begin{flushright}
\end{flushright}
\vspace*{5mm}

\begin{center}
\vspace{1.cm}
{\Large\bf
Exploring  the  conformal transition}\\
\vspace{.4cm}
 {\Large\bf from above and below}\\
\vspace{1.5cm}
{\large Alex Pomarol$^{a,b}$ and Lindber Salas$^{a,b}$}\\
\vspace{.6cm}
$^a${\it {IFAE and BIST,  Universitat Aut{\`o}noma de Barcelona,
08193 Bellaterra, Barcelona}}\\
$^b${\it {Dept. de F\'isica,  Universitat Aut{\`o}noma de Barcelona,
08193 Bellaterra, Barcelona}}\\
\vspace{.4cm}
\end{center}

\vspace{1cm}
\begin{abstract}
We consider conformal transitions arising from the merging of IR and UV fixed points, expected to occur in QCD with a large enough number of flavors. We study the smoothness of  physical quantities  across this transition, being mostly determined by the  logarithmic breaking of conformal invariance. We investigate this explicitly using holography where approaching the conformal transition either from outside or inside the conformal window (perturbed  by a  mass term)   is  characterized by  the same dynamics. The  mass  of spin-1 mesons and $F_\pi$  are   shown to be  continuous across the transition, as well as the  dilaton mass. This implies that the lightness of the dilaton cannot be a  consequence of the spontaneous breaking of scale invariance when leaving the conformal window. Our analysis suggests  that  the light scalar observed  in QCD lattice simulations is  a  $q\bar q$ meson that becomes light   since the  $q\bar q$-operator dimension reaches its minimal value.
\end{abstract}

\vfill
\eject
\pagestyle{empty}
\setcounter{page}{1}
\setcounter{footnote}{0}
\pagestyle{plain}
%

\section{Introduction}

A conformal field theory (CFT) can depart  from its IR fixed point 
in various way  as we vary the  parameters of the model.
Either because  the IR fixed point goes to zero, to infinity or it merges  with a UV fixed point.
We are interested in   conformal transitions  characterized by this  third case,
the merging of the IR fixed point with a UV fixed point.

\begin{figure}[t]
\centering
\includegraphics[width=.55\textwidth]{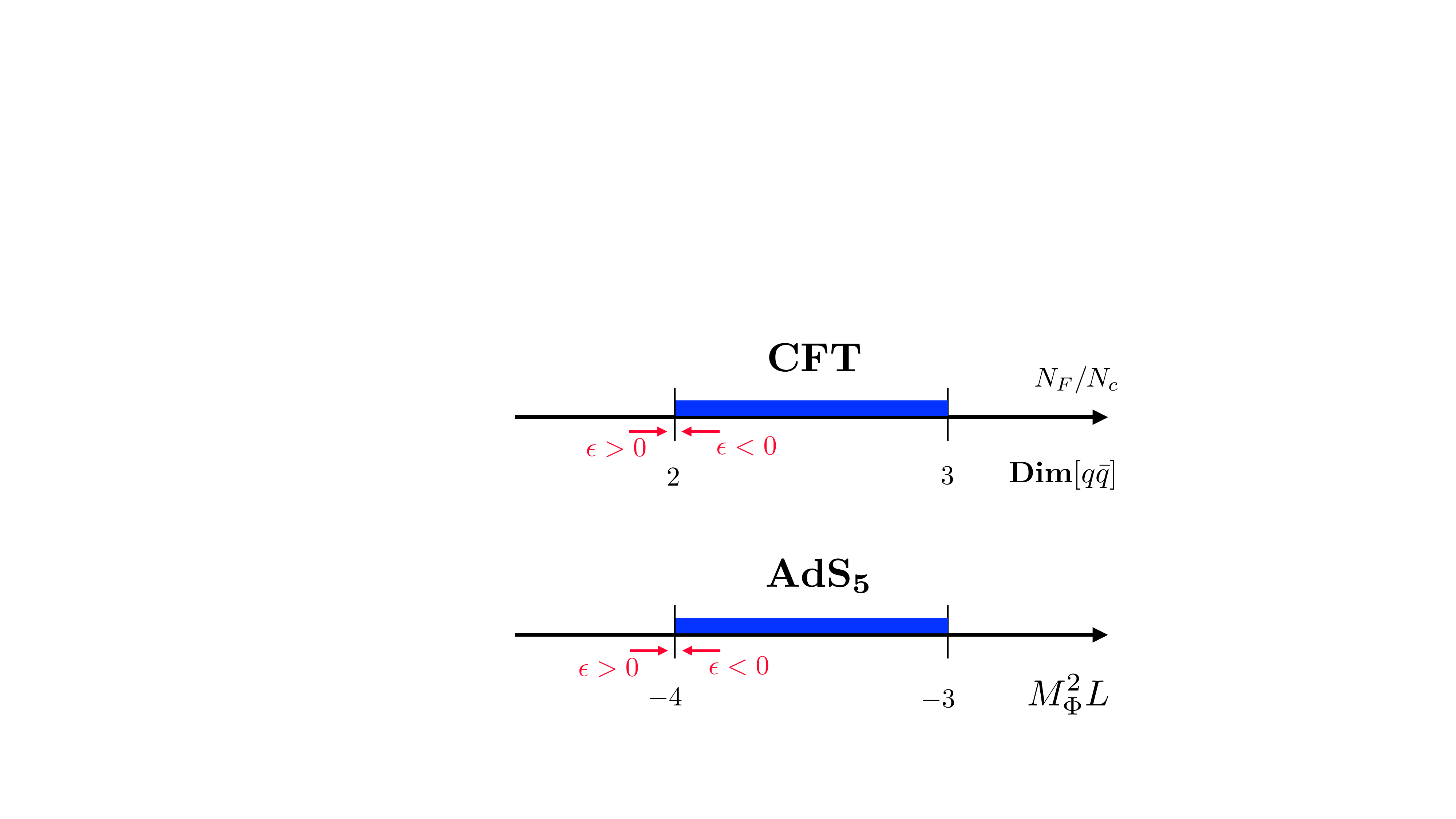}
\caption{\it QCD conformal window and the holographic equivalent.}
\label{cftwindow}
\end{figure}

It has been speculated \cite{Kaplan:2009kr} that this is the case for $SU(N_c)$ gauge theories as QCD at the lower edge of the conformal window --see Fig.~\ref{cftwindow}. 
As we decrease   the number of flavors $N_F$ from the  Banks-Zaks fixed point   at $N_F=\frac{11}{2}N_c$, where QCD enters into  the conformal window, to
some critical value $N_F=N_F^{crit}$,   QCD is expected to loose conformality by an IR-UV fixed point merging.
Interestingly, in the last years lattice simulations  have been providing  abundant  data on 
the properties  of QCD at different values of $N_F$ and quark masses  $M_q$, helping  to better understand this conformal transition
 \cite{Aoki:2016wnc,Brower:2015owo,Appelquist:2016viq,Fodor:2011tu}. 
A particularly intriguing  feature is 
 the presence of a very  light  $0^{++}$ state  when  QCD  is close  to (but outside) the conformal transition.
 It has been speculated that this  state could be a dilaton, the Goldstone associated to the spontaneous breaking of scale invariance.

We will consider CFTs in the  large-$N_c$ limit.
It has been argued in 
 \cite{Pomoni:2008de,Gorbenko:2018ncu}  that when these models 
are  close to the conformal transition, they   must contain  a scalar operator ${\cal O}_\Phi$
whose  dimension gets close to $2$,  becoming  imaginary when leaving the conformal window.
For  QCD, where   in the  large-$N_c$ limit  $N_F^{crit}/N_c\equiv x_{crit}$  becomes a continuous parameter,
 ${\cal O}_\Phi$ corresponds to the $q\bar q$ operator.

We are interested in understanding how the physical quantities change as we move across the conformal transition.\footnote{Effective field theories (EFTs) for a light dilaton have been widely developed  \cite{Golterman:2016lsd,Appelquist:2017wcg}. Nevertheless, these are limited to small $M_q$ values and cannot be used to  
describe the conformal  transition.} 
 Using holography \cite{adscft} we will show how the meson mass spectrum is mostly dictated by  chiral and conformal invariance  and the way this is broken. 
 We will show that spin-1 meson masses and $F_\pi$ are continuous across the  
transition, while the masses of the scalar mesons, $f_0$ and $a_0$, 
show a  jump due to a logarithmic  breaking of  conformal invariance.

The mass of the dilaton, corresponding to a glueball, is also found to be smooth across the transition, implying that  
this must be light at both sides of the conformal window.
This implies that the lightness of the dilaton cannot only be a  consequence of the spontaneous breaking of the conformal symmetry when leaving the conformal window.
We  will  argue that this disfavors this state as the light  $0^{++}$ scalar found in lattice simulations.

On the other hand, we will consider the possibility that 
the lightness of the $0^{++}$ scalar found in lattice simulations is a $q\bar q $ meson 
whose mass is small due to the fact that Dim$[{\cal O}_\Phi]\to 2$ at the conformal edge. 
This is  the  closest value to the unitary bound  Dim$[{\cal O}_\Phi]\geq 1$ where  a scalar is expected to become massless since  the operator ${\cal O}_\Phi$  decouples from the CFT \cite{Grinstein:2008qk}.
We will also understand why the breaking of the chiral symmetry is smaller as we move  inside the conformal window,
 as lattice simulations seem to suggest.

Although some of these properties could also be derived following   4D CFT approaches, as  for example in
 \cite{Cohen:1988sq} or  \cite{Zwicky:2023bzk}, we will see that it is much easier to derive  them using   holography.

\section{Conformal transition by fixed-point merging}
\label{ini}

Following \cite{Kaplan:2009kr}, we will consider
that the  conformal transition occurs  when an IR fixed point  merges with a UV fixed point.
For theories at large-$N_c$, 
it can be shown \cite{Pomoni:2008de,Gorbenko:2018ncu} 
that  close to the transition the  theories must have  a marginal operator ${\cal O}_f$
\be
f(\mu) {\cal O}_f\ \in {\cal L}\,,
\label{marginal}
\ee
whose coupling  $f$ has a beta function given by
\be
\beta_f\simeq \epsilon+(f-f_*)^2\,.
\label{betag}
\ee
For $\epsilon<0$ this beta function has two zeros corresponding to 
an IR and a UV fixed point that  merge to a single fixed point at $\epsilon=0$, that disappears for $\epsilon>0$.
This marks the conformal transition. In QCD we expect  $\epsilon\propto x_{crit}-x$.

Therefore, as we approach the conformal transition ($\epsilon\to 0$), 
the theory can be in two different phases depending on the sign of $\epsilon$:
\begin{itemize}
\item
For   $\epsilon<0$, \eq{betag} gives us
\be
f(\mu)\simeq f_*-\frac{1}{\ln \mu/\mu_0}\,,
\label{frun}
\ee
and $f$  will run  towards the IR, $\mu/\mu_0\to 0$, 
approaching  $f_*$ where the theory becomes conformal, $\beta_f\to 0$. Here $\mu_0$ is an arbitrary scale related with the value of  $f$ at the UV.
\item
For   $\epsilon>0$,  there are no possible zeros for $\beta_f$. In this case, 
we have
\be
f(\mu)\simeq f_*+\sqrt{\epsilon} \tan\left(\sqrt{\epsilon} \ln\frac{\mu}{\mu_0}\right)\,,
\label{frun2}
\ee
and $f(\mu)$ runs slowly when  $f \sim f_*$, behaving almost as a CFT, but it blows up at some IR scale 
$\mu_{\rm IR}$ determined by 
\be
\sqrt{\epsilon} \ln\frac{\mu_{\rm IR}}{\mu_0}\simeq -\frac{\pi}{2}\,.
\label{IRscale}
\ee
Therefore conformality is never reached.
In fact,
the dimension of ${\cal O}_f$,  formally given by
\be
{\rm Dim}[{\cal O}_f]=4+\frac{d\beta_f}{df}\simeq4+2\sqrt{-\epsilon}\,,
\label{complexdim}
\ee
 becomes complex for $\epsilon>0$.
\end{itemize}
These two phases corresponds to the two sides of the conformal  lower edge shown in Fig.~\ref{cftwindow}.

It has also been shown in \cite{Pomoni:2008de,Gorbenko:2018ncu} that   for  theories in  the large-$N_c$ limit,
${\cal O}_f$ must be a double-trace operator, made of the  squared  of a single-trace operator ${\cal O}_\Phi$, i.e.,
${\cal O}_f=|{\cal O}_\Phi|^2$.
Since in the large $N_c$,  ${\rm Dim}[{\cal O}_f]=2{\rm Dim}[{\cal O}_\Phi]$, we have  from \eq{complexdim}
\be
{\rm Dim}[{\cal O}_\Phi]=2+\sqrt{-\epsilon}\,.
\label{complexdim2}
\ee
In QCD, as argued  in \cite{Kaplan:2009kr}, 
 ${\cal O}_\Phi$  is  expected  to be the   operator made of quarks, $q\bar q$, whose dimension will  go from $\sim 3$ 
 when entering the conformal window at the upper edge  
  to 2   at the lower edge (see Fig.~\ref{cftwindow}).
The fact that  ${\cal O}_\Phi$ reaches the lowest  dimension at the conformal transition 
can  explain  the existence of a relative light  scalar meson (with respect to the vector one, $m_\rho$) \cite{ours}.
Indeed, Dim$[{\cal O}_\Phi]$ gets at the edge of the conformal transition  the closest value to  Dim$[{\cal O}_\Phi]=1$ (unitarity bound) at which
the scalar operator decouples from the CFT \cite{Grinstein:2008qk},  becoming then insensitive to the CFT IR scale.

All the above properties of this conformal transition find a beautiful implementation in  holographic models by the use 
of the correspondence (or duality) between strongly-coupled CFT$_4$ (in the large $N_c$ and large 'tHooft coupling\footnote{It has been recently shown \cite{Albert:2022oes,Fernandez:2022kzi}
 that higher spin decouple from low-energy observables, making  holography  a good approach even for models where the 'tHooft coupling is not very large.})
and  weakly-coupled five-dimensional Anti-de-Sitter theories (AdS$_5$) \cite{adscft}.
Operators  in the CFT$_4$ (${\cal O}_\Phi$) correspond to scalar fields in the AdS$_5$  ($\Phi$)
where dimensions and masses are related via the  the AdS/CFT relation \cite{adscft}:
\be
{\rm Dim}[{\cal O}_\Phi]=2+\sqrt{4+M^2_\Phi L^2}\,. 
\label{dic1}
\ee
\eq{dic1} tells us that the conformal transition must occurs when  the 5D scalar  mass
$M^2_\Phi L^2$  becomes smaller than  $-4$ (see Fig.~\ref{cftwindow}).
Indeed, in this case the mass is below the Breitenlohner-Freedman (BF) bound  that determines  the stability of a scalar in AdS$_5$.
For $M^2_\Phi <-4/L^2$ the scalar $\Phi$ becomes  tachyonic, turning  on in the 5D bulk \cite{Kutasov,ours}.\footnote{Alternatively, one could consider  holographic models  of   complex CFTs, as done in \cite{Faedo:2019nxw}.}

\subsection{Probing the  conformal phase  by $M_q {\cal O}_\Phi$}

We are interested in understanding the properties of the mass spectrum at the two sides of the conformal edge.
Since the spectrum in the conformal phase is continuous,  we will perturb  the theory by adding to the Lagrangian the term
\be
\Delta {\cal L}=M_q {\cal O}_\Phi\  \ \ \ (M_q\geq 0)\,,
\label{mqadd}
\ee
that   explicitly breaks  scale invariance.
In QCD this corresponds to add a mass to the quarks, $M_qq\bar q$, that not only breaks  conformal invariance but also  the chiral symmetry; this is  also done in lattice simulations.

A nonzero $M_q$   allows to probe the physical properties of the theory inside the conformal window,
as the mass spectrum  becomes discrete and can be compared with the one at the other side of the edge of the transition. 
All the masses are expected to be proportional  to $M_q^{1/(4-{\rm Dim} [{\cal O}_\Phi])}$, referred as "hyperscaling". 
At the conformal edge where Dim$[{\cal O}_\Phi]\to 2$, we then have
\be
m_i=a_i \sqrt{M_q}\,,
\label{hyperscaling}
\ee
where $a_i$ are parameters that depend on the details of the model.
Obviously, the ratio of masses  is independent of $M_q$. 

The presence of $M_q$ in the conformal theory ($\epsilon<0$) brings also a logarithmic divergence that can be easily understood from
scale invariance \cite{Witten:2001ua}.
Since   dim$[{\cal O}_\Phi]\to 2$,  
 the two-point function in momentum space is given by
\be
\int d^4 x  e^{ip\cdot x }\langle{\cal O}_\Phi ( x){\cal O}_\Phi(0)\rangle 
\sim \int d^4 x  \frac{1}{|x|^4}\sim \ln\Lambda\,. 
\ee
Therefore we expect  a logarithmic breaking of conformal  invariance in 
$\langle{\cal O}_\Phi\rangle$ proportional to  $M_q$. Notice however that $M_q$ does not enter into the log,
so \eq{hyperscaling} is always guaranteed. 

From outside the  conformal window ($\epsilon>0$), the situation is different.
If we add \eq{mqadd} and increase $M_q$ over   the scale  $\mu_{\rm IR}$ defined in \eq{IRscale}, the coupling
$f(\mu)$  stays almost constant  around $f_*$ as in a CFT.  
Therefore in the limit  $M_q\gg \mu_{\rm IR}$ the mass spectrum must smoothly tend to \eq{hyperscaling}.

Before moving to the holographic model, 
we must remark that our analysis using holography shares 
many features of  that in \cite{Cohen:1988sq}
based on the Schwinger-Dyson equation for the 
renormalized  fermion self-energy.
It is also  related to the approach taken in \cite{Zwicky:2023bzk} where the theory is assumed to be conformal deep in the IR.

\section{A  five-dimensional model for the conformal transition}
\label{model5D}

A  holographic model with the properties of described above 
was presented in \cite{ours} (for other models, see \cite{Kutasov,lightdilaton}). 
 This   consists in a    $SU(N_F)_L\otimes SU(N_F)_R$ gauge theory in 5D with a complex scalar $\Phi$ 
transforming as a  $({\bf N_F}$,${\bf\overline  N_F})$.\footnote{With respect to \cite{ours},  we are neglecting the $U(1)_B$ gauge sector that does not play any role in the discussion.}
This scalar plays the role of the  $q\bar q$  operator in 4D QCD.
Imposing parity ($L \leftrightarrow  R$),  the action is given by 
\begin{equation}
S_5=\int d^4x\int dz\, \sqrt{g}\,  M_5 \left[\frac{1}{\kappa^2}\left({\cal R}+\Lambda_5\right)+{\cal L}_5\right]\, ,
\label{s5}
\end{equation}
where the Lagrangian is given by
\begin{equation}
{\cal L}_5 = 
-\frac{1}{4}\Tr \left[ L_{MN}L^{MN}+R_{MN}R^{MN} \right]
 +\frac{1}{2} \Tr |D_M\Phi|^2
-V_\Phi(\Phi)\,,
\label{la5}
\end{equation}
with $L_{MN}$, $R_{MN}$  being the field-strength of the $SU(N_F)_L$ and  $SU(N_F)_R$
 gauge bosons respectively,  and the indices  run  over the five dimensions, $M=\{\mu,5\}$.
We parametrize the fields
as $\Phi=\Phi_s+T_a\Phi_a$
 with  Tr$[T_aT_b]=\delta_{ab}$.
 The fields $\Phi_s$ and $\Phi_a$ will respectively transform as singlet and adjoint under $SU(N_F)_V$,
 the remaining symmetry after $\Phi\not=0$ breaks the chiral symmetry.
The  covariant derivative is defined as
\begin{equation}
D_M\Phi=\partial_M \Phi+ig_5L_M\Phi-ig_5\Phi R_M\, ,
\end{equation}
and the potential is given by\footnote{We notice that one can absorb  one coupling into $M_5$, as we will do later --see footnote \ref{foot}.}
\be
V_\Phi(\Phi)=\frac{1}{2}M^2_\Phi\Tr|\Phi|^2+\frac{1}{4}\lambda_1 \Tr|\Phi|^4+\frac{1}{4}\lambda_2 (\Tr|\Phi|^2)^2\, .
\label{pot5D}
\ee
The 5D metric in conformal coordinates is defined as
\begin{equation}
ds^2=a(z)^2\big(\eta_{\mu\nu}dx^\mu dx^\nu-dz^2\big)\, ,
\label{metric}
\end{equation}
where   $\eta_{\mu\nu}=\diag(1, -1, -1, -1)$ and 
$a(z)$ is the warp factor. Before the scalar $\Phi$ turns on, the presence of $\Lambda_5$ leads to  an     AdS$_5$ geometry: 
\begin{equation}
a(z)=\frac{L}{z}\,,
 \label{ads}
\end{equation}
where  $L^2=12/\Lambda_5$ is the squared AdS curvature radius.
The  5D space will be cut off  by an  IR-brane at some point   $z=\zir$ to be  determined dynamically.
Also a UV-boundary at $z=\zuv$ will be needed to regularize the theory.   The limit $\zuv\to 0$ will be taken in a proper way to  provide  finite physical quantities \cite{adscft}.

The dimension of ${\cal O}_\Phi$,   \eq{complexdim2}, is related by \eq{dic1} to the 5D mass of $\Phi$:
\be
M^2_\Phi=-\frac{4+\epsilon}{L^2}\,.
\label{mass5D}
\ee
When $\epsilon<0$ the mass of $\Phi$ is above the BF bound  and $\Phi$ does not turn on, as we will see in Sec.~\ref{312}. Nevertheless, for  $\epsilon>0$ the  mass  is below the BF bound and   $\Phi$  turns on in the 5D bulk \cite{ours},
breaking the conformal and chiral symmetry.
Therefore, as in the strongly-coupled model described in Sec.~\ref{ini}, 
we have two phases separated  by the sign of $\epsilon$  (see Fig.~\ref{cftwindow}):
\begin{itemize}
\item  $\epsilon>0\ \Rightarrow\ $ non-AdS$_5$ (non-CFT$_4$) phase.   
\item  $\epsilon<0 \ \Rightarrow\ $  AdS$_5$  (CFT$_4$) phase.
\end{itemize}
The presence of  the IR-brane add extra parameters to the theory as
$\Phi$ might also have a potential on the IR-boundary. Following the EFT criteria of \cite{ours}, we have 
\begin{equation}
{\cal L}_{\rm IR} = -a^4 \tilde V_b(\Phi)\big|_{\zir} ,\qquad \tilde V_{b}(\Phi)=\frac{\Lambda_4}{\kappa^2}+\frac{1}{2}m^2_b\Tr|\Phi|^2\,.
\label{pot4D}
\end{equation}

\subsection{The $\phi(z)$ profile}

The conformal and the  chiral symmetry breaking 
$SU(N_F)_L\otimes SU(N_F)_R\to SU(N_F)_V$
is trigger  by  a nonzero  profile for  $\phi={\rm Tr}|\Phi|$.
From \eq{la5},  the equation of motion for $\phi$ is determined to be
\be
-\frac{1}{a^5} \left( \partial_5\, a^3\partial_5-a^3 \partial^\mu\partial_\mu \right) \phi+M^2_\Phi\, \phi+ \lambda\, \phi^3=0\,,
  \ee
where  $\lambda\equiv \lambda_1+N_F\lambda_2$ and the warp factor is determined by the Einstein equations \cite{ours}:
\be
-\frac{\dot a}{a^2}=\sqrt{\frac{1}{L^2}+\frac{\hat\kappa^2L^2}{12}\Big(\frac{\dot{\phi}^2}{2a^2}-V(\phi)\Big)}\,,
\label{wf}
\ee
with $\dot \phi\equiv\partial_z \phi$ and $\hat \kappa^2$ being the 5D gravitational strength. 
The boundary conditions are the following.
At the UV-boundary we fix
\be
\left. L\phi\right|_{\zuv}=\zuv^2 M_q\,.
\label{bcuv}
\ee
$M_q$ plays the role of the quark mass in the dual gauge theory, \eq{mqadd}. It has dimension
\be
{\rm Dim}[M_q]=4-{\rm Dim}[{\cal O}_\Phi]\,,
\ee
so at the conformal edge we have   Dim$[M_q]\to 2$. 
For $M_q\not=0$,  $\phi$  turns  on  independently of the sign of $\epsilon$,
and the conformal and chiral symmetry are broken inside the 5D bulk. 
For $M_q=0$, the field $\phi(z)$ gets a nonzero profile
only when we are outside the conformal window ($\epsilon>0$), as we will explicitly see later.
At the IR-brane  we must impose the boundary condition determined by the model. 
We have \cite{ours}
\be 
\left. \left(\frac{M_5}{a} \dot\phi +\partial_\phi V_b\right)\right|_{\zir}=0\  ,\ \ \ \ 
\left.  \left(-\frac{6 M_5}{\hat \kappa^2L^2}\frac{\dot a}{a^2}+ V_b\right)\right|_{\zir}=0\,.
\label{bcir}
\ee
The first one is the IR condition  for $\phi$, while the second one is the junction condition that determines the position
of the IR-brane $\zir$.

Although we will present in the next section  results with no approximations, it is instructive to consider the case in which the profile of $\phi$ can be solved analytically. 
For this, we will take the approximation  that $\phi$ is small such that the quartic term can be neglected as well as  the  feedback of $\phi$ on the metric, i.e., the 5D space is  AdS$_5$. 
\eq{bcir}   reduces in this case to
\be
\dot\phi(\zir) =-\frac{m^2_bL}{\zir M_5}\phi_{\rm IR}\ ,\ \ \  \phi(\zir) =\phi_{\rm IR}\,,
\label{bcirs}
\ee
where   we have introduced the parameter $\phi_{\rm IR}$  related to other parameters of the  model (a combination of $\hat \kappa^2$, $ m^2_b$, $\Lambda_4$ and the sign of $\lambda$ \cite{ours}).\footnote{\label{foot}
By field redefinitions we can absorb $|\lambda|$ in other parameters.
Therefore, with no loss of generality, we can consider $\lambda=\pm 1$.}
We will restrict to $m^2_b>-2M_5/L$ that guarantees that the conformal  symmetry is not broken  by the IR-boundary potential.

\subsubsection{Towards the conformal transition from outside ($\epsilon>0$) }
\label{epsp}
With the above approximations, we can solve $\phi$ analytically. For $\epsilon>0$, the two solutions
are $z^{2\pm i\sqrt{\epsilon}}$ that we can write as
\be
\phi(z)=\frac{A}{L^3} z^2 \sin\left(\sqrt{\epsilon} \ln\frac{z}{\zuv}+\beta\right)\,,
\label{eqtachyon}
\ee
where $A$ and  $\beta$ are dimensionless constants to be determined by the boundary conditions. 
From \eq{bcuv} and \eq{bcir}, we get
\be
M_q=\frac{A}{L^2}\sin \beta\,,
\label{mq}
\ee
and
\be
\tan \left(\sqrt{\epsilon} \ln\frac{\zir}{\zuv}+\beta\right)=-\frac{\sqrt{\epsilon}}{\tilde m^2_b}\,,
\label{bcnew}
\ee
where $\tilde m^2_b\equiv 2+m^2_bL/M_5$.
In the limit $\epsilon\to 0$, \eq{bcnew} tells us that  $\sin(\sqrt{\epsilon} \ln \zir/\zuv+\beta)\to \sqrt{\epsilon}$.
Therefore expanding \eq{bcnew} around $\zc$ determined by\footnote{The solution for   $n=1$ corresponds to a  global minimum, while  $n>1$ just gives   local minima
(corresponding to a surviving discrete conformal invariance).}
\be
\sqrt{\epsilon} \ln\frac{\zc}{\zuv}+\beta= n\pi\ ,\ \  n=1,2,...\,,
\label{zirc}
\ee
we get 
\be
\ln\frac{\zir}{\zc}=-\frac{1}{\tilde m^2_b}<0\,. 
\label{new}
\ee
Notice that  the limit $\epsilon\to 0$ must be taken with $\zuv\to 0$ according to  \eq{zirc}. 
Using \eq{bcirs} and  \eq{new} we finally get 
\be
\phi(z)=-\phi_{\rm IR}\, \tilde m_b^2 \left(\frac{z}{\zir}\right)^2  \ln\frac{z}{a\,\zir}\,,
\label{eqtachyon2}
\ee
where $a={\rm Exp}[1/\tilde m_b^2]$.
It is interesting to remark that \eq{eqtachyon2} is valid for any value of $M_q$.
In the particular case $M_q=0$, we have from \eq{mq}  that $\beta=0$ but $A\not=0$,  corresponding to a spontaneous breaking of the conformal and chiral symmetry. 

The above   is in accordance with the discussion in Sec.~\ref{ini} where for $\epsilon>0$ it was shown 
that  $f(\mu)$   runs as \eq{frun2} and diverges at the scale $\mu_{\rm IR}\sim 1/\zc$ for $\mu_0\sim {\rm Exp}[-\pi/(2\sqrt{\epsilon})]/\zuv$.

\subsubsection{Towards the conformal transition from inside ($\epsilon<0$)}
\label{312}

Let us now consider the solution of $\phi$ from the other side of the conformal edge, $\epsilon<0$. In this case
the two possible solutions are $z^{2\pm \sqrt{|\epsilon}|}$ that in  the limit  $\epsilon\to 0$ leads to   $z^2$ and $z^2\ln z$.
We can then write  the most general solution as
\be
\phi=\frac{z^2}{L^3}{\hat A} \ln\frac{z}{z_0}\,,
\ee
where ${\hat A}$ and $z_0$ are the two parameters to be fixed by the boundary conditions.
\eq{bcuv} gives
\be
M_q=\frac{{\hat A}}{L^2}\ln\frac{\zuv}{z_0}\,.
\label{mq2}
\ee
From the IR-brane boundary conditions we get, similarly to \eq{new},  
\be
\ln\frac{\zir}{z_0}=-\frac{1}{\tilde m^2_b}<0\,,
\label{new2}
\ee
that leads exactly to \eq{eqtachyon2}. 
We notice however 
an important difference in this   case  with respect to the  $\epsilon>0$ case. 
From \eq{mq2}  we have that  $M_q=0$ requires  $z_0\to \zuv$, but this is  incompatible with  \eq{new2}.
In other words, there is no nonzero solution for $\phi$ when $M_q=0$. 
As expected, the model flows to the  conformal  phase for $M_q=0$.

We can then conclude from the above analysis that approaching the conformal  transition $\epsilon\to 0$ from inside 
the conformal window ($\epsilon<0$) or outside ($\epsilon>0$)  gives the same profile for $\phi$  (\eq{eqtachyon2}) 
and, as a consequence,  the spontaneous conformal  and chiral breaking driven by $\phi$ 
at the IR have to be felt equally in both sides of the transition,   independently of the value of $M_q\not=0$.

\begin{figure}[t]
\centering
\includegraphics[width=0.7\textwidth]{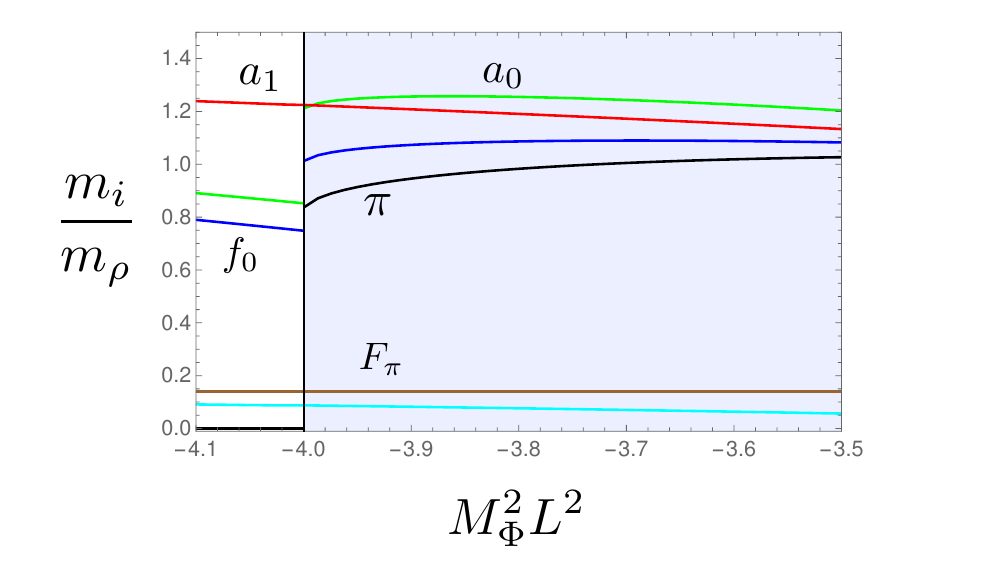}
\caption{\it Mass spectrum of mesons, normalized to $m_\rho$,  as a function of the 5D scalar mass or equivalently $Dim[q\bar q]$  defined in   \eq{dic1}. 
The values of the model are given in \eq{values}.  We vary  the value of $\phi_{\rm IR}$ to keep $F_\pi/m_\rho$ fixed. The sky-blue line corresponds to the dilaton/radion (glueball). 
For $M^2_\Phi L^2<-4$ we have $M_q=0$, while for $M^2_\Phi L^2>-4$ we have $M_q\not=0$.}
\label{massesvsmphi}
\end{figure}

\section{Mass spectrum}

Let us discuss here what differences we expect  in the mass spectrum of the theory  when we approach the 
conformal edge from inside or outside the conformal window. 
We will back up our arguments with the mass spectrum  calculated in the holographic model with no approximations.
For the numerical analysis we  will take the  benchmark values
\be
\phi_{\rm IR}=1\,,\ \ \lambda=1\,,\  \ N_F\lambda_2=-2\,, \ \ \tilde m^2_b=1\,,\ \hat \kappa=1\,,\ \  g_5=1.52\,.
\label{values}
\ee 
The mass spectrum would change by varying these values, but the qualitative picture will be the same. For other values of the parameter space  see \cite{ours}. 

\subsection{Spin-1 states}

It is clear that  vector states, coming from the Kaluza-Klein (KK) decomposition of $V_M=(L_M+R_M)/\sqrt{2}$ are not much affected by the scalar $\phi(z)$ since they do not couple to it. They can only notice a nonzero $\phi$ from the feedback of  this on the metric.  This clearly affects  the KK spectrum, but this is expected to be quite universal for the different type of states. 
For this reason, we will use the mass of the lightest vector state, the $\rho$, to normalize the other masses.

The  axial-vector  $A_M=(L_M-R_M)/\sqrt{2}$ couple to $\phi$ through the covariant derivative, and therefore
a nonzero $\phi$ splits the masses of the KK  of  $A_M$    from those of  $V_M$.
Since $\phi$ has the same profile at both sides of the conformal edge independently  of $M_q$, as shown in \eq{eqtachyon2}, we expect   the masses of  the axial-vectors to be  smooth across the transition.
Similarly for $F_\pi$,  defined as the axial-vector two-point correlator at zero momentum \cite{ours}, we  expect this quantity to be independent of $M_q$ and smooth across the transition.

\begin{figure}[t]
\centering
\includegraphics[width=0.7\textwidth]{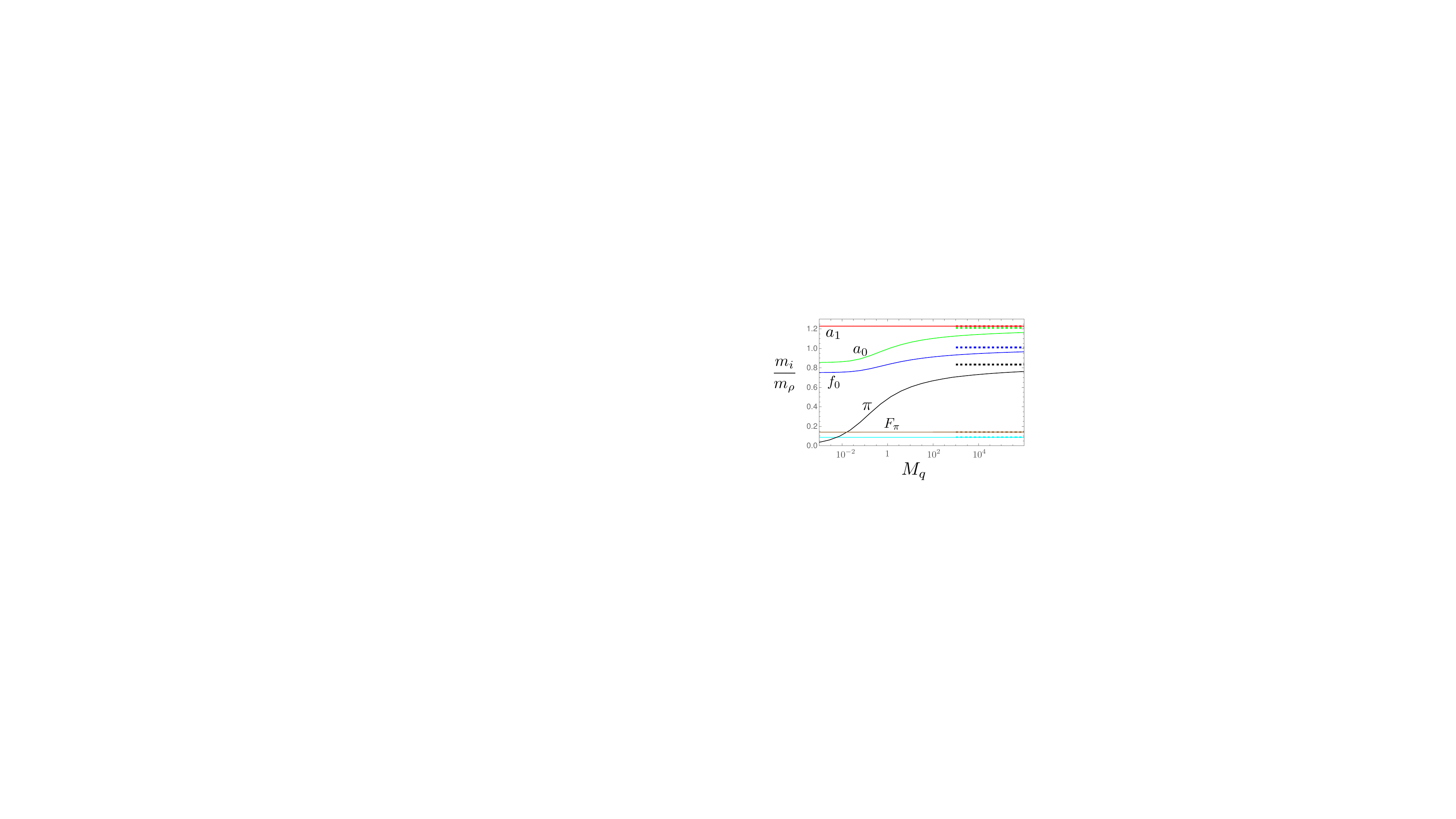}
\caption{\it Mass spectrum of mesons outside the conformal window $\epsilon>0$, normalized the $m_\rho$, as a function of $M_q$. In dashed-lines the values for $\epsilon<0$. The sky-blue line corresponds to the dilaton/radion (glueball).}
\label{massesvsmq}
\end{figure}

In Fig.~\ref{massesvsmphi} we show the mass of the lightest axial-vector, $a_1$
as well as $F_\pi$  normalized to $m_\rho$. We indeed see that these values are smooth across the transition.  
To show that these physical quantities are  also independent of $M_q$, we 
 plot in  Fig.~\ref{massesvsmq} the predictions of the holographic model as a function of $M_q$
 for  $\epsilon>0$.
 We remark again that this is
obvious for $\epsilon<0$ (hyperscaling) but not for   $\epsilon>0$. 
 Indeed we see in Fig.~\ref{massesvsmq} that $a_1$ and $F_\pi$ {(normalized to $m_\rho$)} do not vary as we move $M_q$. 
 
\subsection{The dilaton/radion}

It has been claimed that  approaching the conformal edge from below, the theory could have a light scalar, the dilaton,
associated to the spontaneous breaking of the conformal invariance.
This has been partly supported by  lattice results \cite{Aoki:2016wnc,Brower:2015owo,Appelquist:2016viq}
that have seen a $0^{++}$ state below the $\rho$ mass for values of $N_F$ where one expects to be
outside (but close to) the conformal window.
In \cite{ours} it was shown that in holographic models the radion, that corresponds to the dilaton in the 4D dual theory,  was the lightest state of the spectrum, although not parametrically lighter than the others (see also \cite{Kutasov,lightdilaton}).
This state arises  from the KK decomposition of the AdS$_5$ gravitons, and therefore should be considered  a glueball state (it mixes with the mesons from $\Phi$ but the mixing comes out to be small \cite{ours}).

Since we showed that the profile of $\phi$ is the same at both sides of the conformal transition, we must also expect
a light radion/dilaton   inside the conformal window (for $M_q\not=0$) as long as we are close to the lower edge.
We find that this is indeed the case. As it can be appreciated
in Fig.~\ref{massesvsmphi} the radion/dilaton mass is smooth across the transition and is also light
 inside the conformal window. 
 The reason for this lightness can be found in \cite{ours} since the argument given there can also be applied at the other 
side of the  transition (for $\epsilon<0$).
This  surprising result  shows that this light dilaton has nothing to do with the spontaneous breaking
of the conformal symmetry.

Another important property of the radion/dilaton is that its mass (normalized to $m_\rho$) is practically  independent  of $M_q$, 
since the profile of $\phi$, given in \eq{eqtachyon2},  is the same for any value of $M_q$.
This is   shown in Fig.~\ref{massesvsmq} by a sky-blue line.

\subsection{Scalar mesons}

Let us now analyze the mass spectrum of the  fluctuations of $\Phi$ corresponding  to $q\bar q$ mesons.
Although they can in principle mix with the glueballs, we find that in our holographic model this mixing is  small.
Let us start with the radial excitations and consider later the angular fluctuations corresponding to the Goldstones. 

\subsubsection{Radial fluctuations: $0^{++}$ states}

To understand the dependence on the sign of $\epsilon$, we will first calculate the 
scalar-scalar correlator on the AdS$_5$ boundary at $z=\zuv\to 0$. 
To obtain analytical expressions, we will work in the approximation where the quartic couplings and the feedback on the metric are neglected. 
At this level the  singlet and adjoint scalars are degenerate. 
Following \cite{DaRold:2005mxj}, we have (neglecting an overall factor)
\be
\Pi_S(p)= M_5 L \left[{2}+{\zuv}
\f{\partial_z {J}_{\sqrt{-\epsilon}}(ip z)+b(p)\partial_z {Y}_{\sqrt{-\epsilon}}(ip z)}{{J}_{\sqrt{-\epsilon}}(ip z)+b(p){Y}_{\sqrt{-\epsilon}}(ip z)}
\right]_{z=\zuv},
\label{correlator}
\ee
where $J_n$ and $Y_n$ are Bessel functions of order $n$ and $p$ is the Euclidean momentum, and
\be
b(p)=-\f{z \partial_z{J}_{\sqrt{-\epsilon}}(ipz)+\tilde m_b^2{J}_{\sqrt{-\epsilon}}(ipz)}{z \partial_z{Y}_{\sqrt{-\epsilon}}(ipz)+\tilde m_b^2{Y}_{\sqrt{-\epsilon}}(ipz)}\Bigg|_{z=\zir}\,.
\ee
\begin{itemize}
\item $ \epsilon<0$:
Let us take the limit $\epsilon\to 0$ from inside the conformal window with $M_q\not=0$ such that  $\zir$ is fixed at some finite value. 
The scalar-scalar correlator \eq{correlator}  defined at  $\zuv\to 0$  simplifies to
\be
\label{ABF3}
\Pi_S(p)=  M_5 L \left[{2}+
\f{2b(p)}{\pi+2b(p)(\gamma+\ln(ip\zuv/2))}\right]+\cdots\,,
\ee
where 
\be
b(p)=-\f{ip\zir{J}_{1}(ip\zir)-\tilde m_b^2{J}_{0}(ip\zir)}
{ip\zir{Y}_{1}(ip\zir)-\tilde m_b^2{Y}_{0}(ip\zir)}\,.
\ee
For large momentum $p\zir\gg 1$, \eq{ABF3} gives
\be
\label{ABF3.1}
\Pi_S(p)\simeq  M_5 L \left[{2}+
\f{1}{\gamma+\ln(p\zuv/2)}\right]\,.
\ee
The origin of the logarithm is expected from the discussion in  Sec.~\ref{ini}.
The theory contains the marginal term $f(\mu) \Tr[{\cal O}_\Phi{\cal O}_\Phi]$ 
where $f$ runs according to \eq{frun}.
For $p\zuv\to 0$ the log-dependent term goes to zero and the 
theory enters into the conformal regime.

Therefore the mass spectrum of the scalars are not sensitive to $\ln \zuv$ terms.
Using the fact that a scalar-scalar correlator   in a large-$N_c$ theory
can also be written   as a sum over infinitely narrow resonances
\begin{align} \label{ABF4}
\Pi_S(p)=\sum_{i=1}^\infty\frac{F^2_{S_i}m^2_{S_i}}{p^2+m^2_{S_i}}\,,
\end{align}
we can obtain the mass spectrum by looking at the poles of \eq{ABF3}.
Taking  $\tilde m^2_b=1$ we find that the lightest resonance, that we named as in QCD $f_0$, is given by 
$ m_{f_0}\simeq {1.26}/{\zir}$,
 much lighter that the $\rho$ meson mass that in the approximation that we are taking here is $m_\rho\simeq 2.4/\zir$. 
 In fact, we find that  $m_{f_0}/m_\rho<1$ is true for any value of  $\tilde m^2_b\geq 0$.

The reason of why the lightest scalar meson  is  lighter than the $\rho$ 
is tied to the fact that the mass of $M^2_\Phi$ is
taking at the conformal edge the lowest possible value ($M^2_\Phi L^2\to -4$) \cite{ours}.
 We can see this analytically by taking the limit $m_{S_i}\zir\gg 1$.
We find 
\be
m_{S_i}\simeq (i-\frac{3}{4}+\frac{\sqrt{-\epsilon}}{2})\frac{\pi}{\zir}\ ,\ \ \  i=1,2,\cdots\,,
\label{massm}
\ee
that shows that the lightest state mass minimizes for $\epsilon=0$. 
As we already said, this can also be understood from the CFT point of view. Close to the conformal lower edge  the dimension of ${\cal O}_\Phi=q\bar q$ takes the closest value to the unitarity bound (Dim$[q\bar q]>1$) 
where a scalar is expected to become massless
since  ${\cal O}_\Phi$  decouples from the CFT \cite{Grinstein:2008qk}.

\item {\bf $\epsilon>0$:}
Let us now consider the limit $\epsilon \rightarrow 0$ from outside the conformal window.
We recall that we have to take this limit such that \eq{zirc} is kept fixed.
This leads to
\be
\Pi_S(p)\simeq  M_5 L \left[2
+\f{2b(p)}{\pi+2b(p)(\gamma+\ln(ip \zc/2)+\tilde M_q)}\right]\,,
\label{BF4}
\ee
where $\tilde M_q=M_q\zir^2/(L\phi_{\rm IR}\tilde m^2_b)$.
Notice that it is now $\zc\sim \zir$ that regulate the logarithm and not $\zuv$ as in the $\epsilon<0$ case.
This means that the  logarithmic breaking of conformal invariance  remains at low-energies.
For large values of $M_q$ however this term tends  to zero and \eq{BF4} approaches  \eq{ABF3}.
This was  expected from the discussion  in Sec.~\ref{ini}:
the presence of a large $M_q$ sets a mass gap to the theory  much larger than the scale 
$\zc$;  the IR flow  "stops" when the theory  is (almost) a CFT, much before the coupling $f(\mu)$ blows up. 
The theory  in this case has to have the same behavior as  that for $\epsilon<0$.

The scalar meson masses are determined by  the poles of \eq{BF4}. We  get
\be\label{BF5}
b(m_{S_i})=-\f{\pi/2}{\gamma+\ln(m_{S_i}\zc/2)+\tilde M_q}\,.
\ee
Taking the approximation $m_{S_i}\zir\gg 1$ and using \eq{new}, we obtain
\be
\label{BF8}
m_{S_i}\simeq 
(i-\f{3}{4})\frac{\pi}{\zir}- \f{\pi/2}{\gamma+\ln({m_{S_i}}\zir/2)+\frac{1}{\tilde m^2_b}+\tilde M_q}\, \frac{1}{\zir} \ ,\ \ \ i=1,2,\cdots\,.
\ee
Notice that the masses are sensitive to a logarithmic conformal breaking (there is no a simple scaling with $1/\zir$),  different   from the case $\epsilon<0$.
Nevertheless, as expected,  \eq{BF8}  tends to \eq{massm} for $M_q\to \infty$. 
\end{itemize}
The numerical results (with no approximations) for the masses of  the lightest scalar singlet ($f_0$) and adjoint ($a_0$)
are shown in Fig.~\ref{massesvsmphi}.
The masses are different at the two sides of the conformal transition due to the log terms discussed above. 
The dependence on $M_q$ for $\epsilon>0$ is shown  in Fig.~\ref{massesvsmq}.
As $M_q$ grows, the masses tend to the values for $\epsilon<0$ (hyperscaling case), shown as  dashed lines.

\subsubsection{Angular fluctuations: the pions}
Let us finally comment  on  the angular  fluctuations of $\Phi$, the pions. These are the only states sensitive to the 
UV-boundary condition and therefore the ones that clearly distinguish among the two limits towards the conformal edge.
As explained above, for $\epsilon>0$ and $M_q=0$ the model shows spontaneous chiral symmetry breaking
and we expect the pions to be massless.  On the other side, $\epsilon<0$, the chiral breaking is explicit (UV driven by $M_q$) and the pions should have a mass as large as the other mesons, following hyperscaling \eq{hyperscaling}. 
This is shown in   Fig.~\ref{massesvsmq}.  It is interesting to remark that we find   $m_\pi< m_{f_0}$ for any value of $M_q$ and for any value of the parameters of the model.

\subsection{Moving further inside  the conformal window}

As  $M^2_\Phi$ moves from  $-4$ to $-3$, corresponding to 
an increase of Dim$[q\bar q]$ from $2$ to $3$, we are  getting further inside the conformal window (see Fig.~\ref{cftwindow}). We expect the scalar mesons to become heavier since we are moving away from the unitarity bound (Dim$[q\bar q]>1$).

On the other hand, the explicit chiral breaking is dictated by $M_q$  whose dimension moves from $2$ to $1$.
As a consequence,
the  profile of $\phi$, that  grows as 
\be
\phi\sim M_q\, z^{{\rm Dim}[M_q]}\,,
\ee
becomes flatter and spreads more into the AdS$_5$ space.
 To keep $F_\pi/m_\rho$ constant,
the flatter the $\phi$ profile, the smaller  $\phi_{\rm IR}$ must be. 
This implies that the effect of $\phi$  on the IR ($z\sim \zir$) becomes relatively weaker.
Since the spectrum of resonances is determined by the fields at $z\sim \zir$,
 we expect that they will notice less the chiral breaking driven by $\phi$. 
This is indeed seen in Fig.~\ref{massesvsmphi} where, as $M^2_\Phi$ moves towards $-3$, 
we have $m_{a_1}\to m_{\rho}$ and
$m_{a_0}\to m_{f_0}\to m_{\pi}$. This behavior is also observed in lattice simulations as we will see below.

\section{Comparison with lattice simulations}

\begin{figure}[t]
\centering
\includegraphics[width=0.45\textwidth]{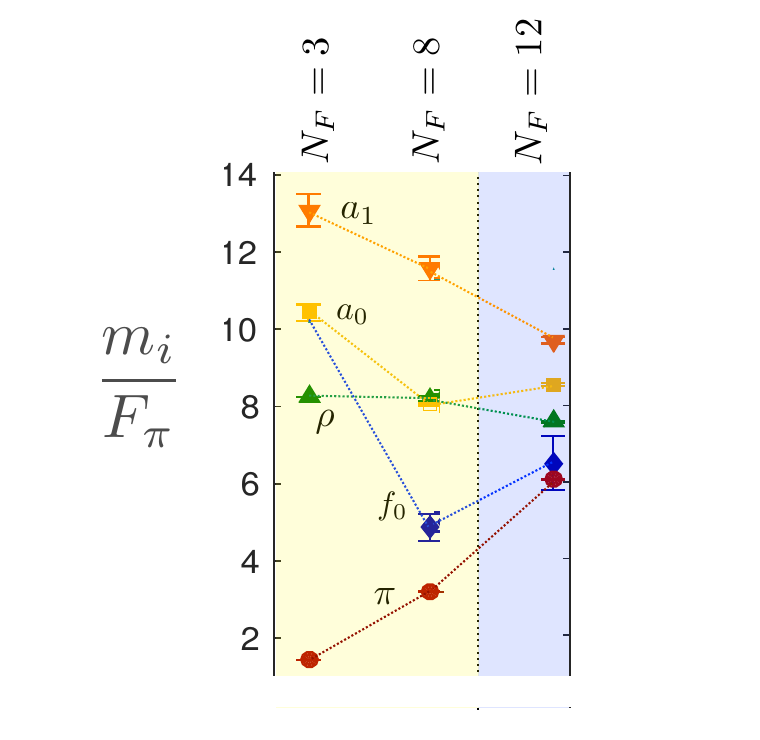}
\caption{\it  Lattice results from \cite{Brower:2015owo}  for the  QCD meson masses normalized to $F_\pi$ for different values of $N_F$. See \cite{Brower:2015owo} for the values of the quark masses.}
\label{lattice}
\end{figure}

There are several  lattice simulations of $SU(3)$ QCD at large values of $N_F$.
In  \cite{Aoki:2016wnc,Appelquist:2016viq}  lattice simulations for $N_F=8$ were provided, while 
results for $N_F=12$ were given in  \cite{Fodor:2011tu}.
  In   \cite{Brower:2015owo}  the value of $N_F$ was  effectively made to vary from $8$ to $12$
by varying the quark masses.
In Fig.~\ref{lattice} we show the  results of  \cite{Brower:2015owo}  for $N_F=3,8$ and $12$.

It is highly supported  that QCD with $N_F=8$ flavors is outside  the conformal window,
while it is inside for  $N_F= 12$. 
The main indication comes from the pion mass that it is seen to  go to zero with $M_q$  for $N_F=8$ and shows hyperscaling for $N_F= 12$ \cite{Brower:2015owo}, as it can be appreciated  in Fig.~\ref{lattice}.
Therefore we can compare our  results inside and outside the conformal window 
with those of lattice for $N_F=8$ and $N_F=12$ respectively.

Fig.~\ref{lattice} shows the following general features:
\begin{itemize}
\item
The scalars are  the lightest states for $N_F=8$ where we expect QCD to be outside (but close) to the conformal window.
\item 
The chiral-breaking mass splittings   diminish as $N_F$ increases and we move inside the conformal window
($m_{a_1}\to m_\rho$, $m_{a_0}\to m_{f_0}\to m_\pi$).
\end{itemize}
These properties are quite close to the ones derived in our holographic model.
Increasing $N_F$  is equivalent to increasing  $M^2_\Phi$ in holographic models 
and    Fig.~\ref{massesvsmphi} shows that  chiral breaking effects become weaker. 
Also the scalar $f_0$ is predicted to be light close to the conformal transition.
Nevertheless, our holographic model also predicts a light
 radion/dilaton that  was advocated in \cite{ours}  to be associated with the lightest scalar seen in lattice simulation.
Nevertheless, the radion/dilaton is predicted to be mostly a glueball whose mass is smooth as we cross the 
conformal transition (sky-blue line in  Fig.~\ref{massesvsmphi}).
This is in contradiction with  lattice results that  seem to suggest that the lightest $0^{++}$ scalar
 is a $q\bar q$ meson, not a glueball, since its mass tends to $m_\pi$ for large $N_f$. 
The  light glueball present in our holographic model  could be then a feature of  the simple   IR-brane setup 
that might be absent in more realistic holographic models.  We leave this investigation for the future.

\begin{figure}[t]
\centering
\includegraphics[width=0.6\textwidth]{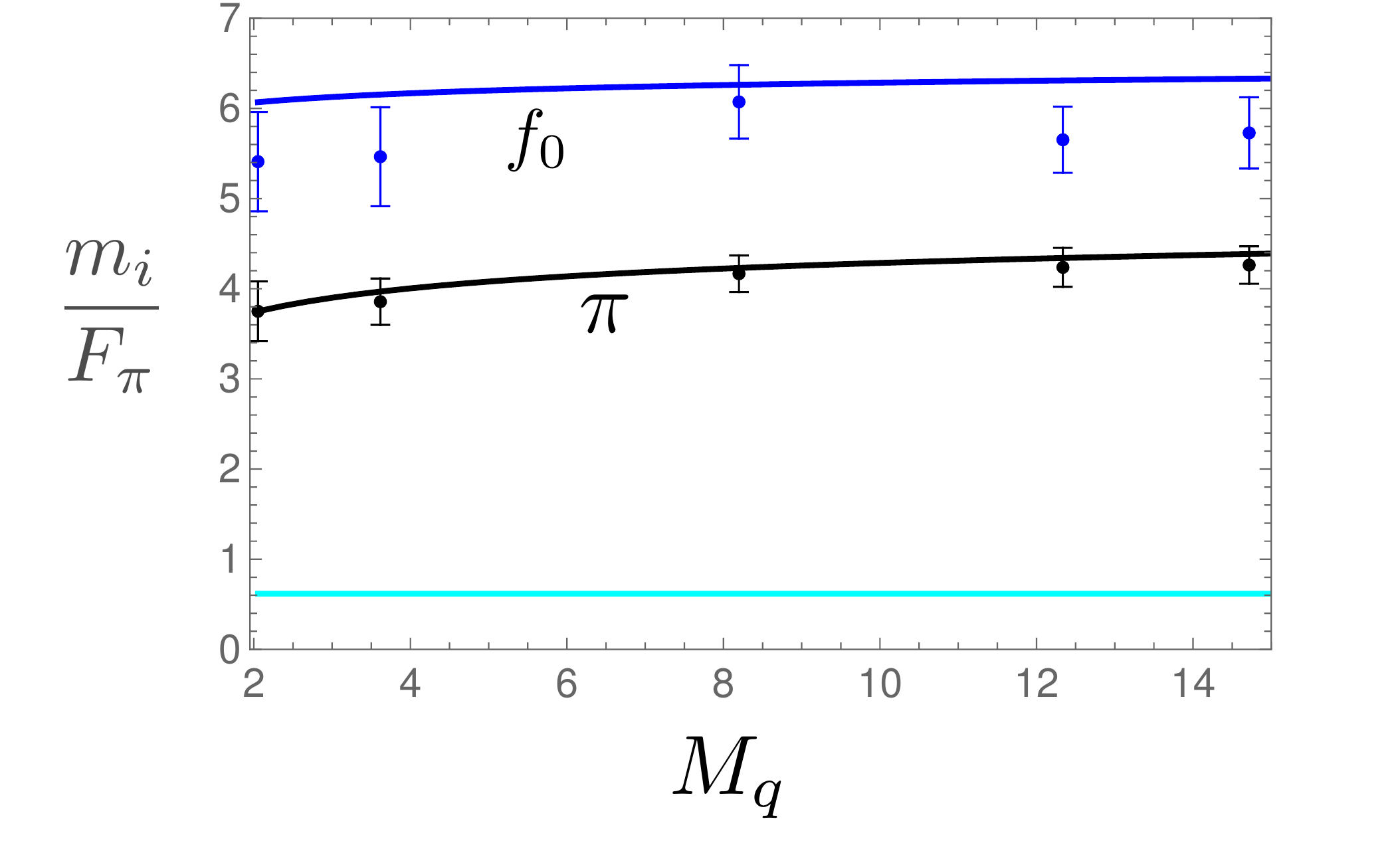}
\caption{\it  Predictions for meson masses normalized to  $F_\pi$ as a function of $M_q$. The sky-blue line corresponds to the dilaton/radion (glueball). Data points are  lattice results  from \cite{LSD:2023uzj}.}
\label{latticecomparison}
\end{figure}

Another important feature derived from our analysis is the dependence of the meson masses with $M_q$, shown in Fig.~\ref{massesvsmq}.
This property might offer an additional hint to the  nature of the light scalar: $q\bar q$ meson if its mass has a $M_q$ dependence 
(see \eq{BF8}), or a glueball if not.
In Fig.~\ref{latticecomparison}
we show our predictions for the masses of $f_0$, $\pi$ and glueball {(normalized to $F_\pi$)} as a function of $M_q$,
and compare them with the lattice predictions of  \cite{LSD:2023uzj}.\footnote{
We have fitted  the lowest value of $m_\pi/F_\pi$  in \cite{LSD:2023uzj} to our prediction  in order to 
normalize our $M_q$ with that in \cite{LSD:2023uzj}.}
Unfortunately, at present lattice results are not  accurate enough to clearly distinguish any dependence with $M_q$, with the exception of $m_\pi$. 
Fig.~\ref{latticecomparison}  however seems to slightly favor   $f_0$ as the lattice $0^{++}$ state in front of a  glueball  state. 

\section{Conclusions}

We have analyzed how the physical properties of a system change 
when  approaches a conformal transition   from both sides of the conformal edge.
The derived properties seem to be generic for large-$N_c$ models where the conformal transition occurs
by the merging of an IR and a UV fixed point.
We have obtained the following features:
\begin{itemize}
\item
The dilaton (mostly a glueball) is light at both sides of the conformal transition, showing that has nothing to do with
the spontaneous breaking of conformal invariance. 
This is very different from the  pion  that is massless outside the conformal window but massive inside due  respectively to the spontaneous and explicit breaking of the chiral symmetry.
\item
The   scalar meson $f_0$ (mostly a $\bar qq$ state) is light  close to the conformal edge, being
lighter when approaching it from outside the conformal window than from inside --see Fig.~\ref{massesvsmq}.
Its mass shows a dependence with $M_q$ predicted as in  \eq{BF8} that can be parametrized as
\be
\frac{m_{f_0}}{m_\rho}= \frac{a_{f_0}}{a_{\rho}} -\frac{\alpha_1}{\alpha_2 \ln\frac{m_{f_0}}{m_\rho}+ M_q}\,,
\ee
where $a_i$ are the hyperscaling values (\eq{hyperscaling}) and $\alpha_i$ are constants. 
\item
Our model predicts $m_\pi< m_{f_0}$  inside the conformal window.
\item
Spin-1 meson masses  as well as  $F_\pi$ are practically  smooth across the conformal transition and independent of $M_q$
as shown in Fig.~\ref{massesvsmq}.
\item
Chiral symmetry breaking effects become smaller as we move further inside the conformal window, i.e.,
 $m_{a_1}\to m_{\rho}$ and $m_{a_0}\to m_{f_0}\to m_{\pi}$.     This is because the dimension of $M_q$, responsible for the chiral and conformal symmetry breaking, decreases.
\end{itemize}
Most of these generic properties seem to be followed by QCD as we increase $N_F$ as lattice results  show  in Fig.~\ref{lattice}.

Models close to the conformal transition can also be useful for physics beyond the SM \cite{ours}.
In particular, if this phase transition occurs in the early universe, a supercooled epoch can generate
interesting  signal  \cite{Baratella:2018pxi,Csaki:2023pwy}. These applications are  left for the future.

\section*{Acknowledgements}
\label{sec:acknowledge}

We would like to thank Oriol Pujolas for collaborating at the early stages of this work.
 LS  also thanks Leandro Da Rold for useful correspondence.
The authors acknowledge  support from the Departament de Recerca i Universitats from Generalitat de Catalunya to the Grup de Recerca "Grup de F\'isica Te\`orica UAB/IFAE" (Codi: 2021 SGR 00649).
This work   has also been  supported by the research grant PID2020-115845GB-I00/AEI/10.13039/501100011033.

\end{document}